\documentclass[prl,showpacs,groupaddress,twocolumn]{revtex4}
\usepackage{graphicx}
\usepackage{amsmath}
\usepackage{bm}
\begin{document} \title{Bogoliubov-like mode in the Tonks-Girardeau Gas}
\author{Igor V. Ovchinnikov}
\author{Daniel Neuhauser}\email{dxn@chem.ucla.edu}
\affiliation{Department of Chemistry and Biochemistry, University of California at Los Angeles, Los Angeles, CA, 90095-1569}
\begin{abstract}
We reformulate 1D boson-fermion duality in path-integral terms. The result is a 1D counterpart of the boson-fermion duality in
the 2D Chern-Simons gauge theory. The theory is consistent and enables, using standard resummation techniques, to obtain the
long-wave-length asymptotics of the collective mode in 1D boson systems at the Tonks-Girardeau regime. The collective mode has
the dispersion of Bogoliubov phonons: $\omega(q)=q \sqrt{\bar\rho U(q)/m}$, where $\bar\rho$ is the bosons density and $U(q)$ is
a Fourier component of the two-body potential.
\end{abstract}
\date{\today}
 \maketitle

1D boson systems attract much attention
\cite{atoms,nanotubes,Ol'shanyi,1DBosonGas,Astra,FermionizedBosons,FermionBosonDuality,ThomasFermi} in light of recent
experiments on cigar-shaped atomic traps \cite{atoms} and gases exposed to linear carbon nanotubes \cite{nanotubes}. The history
of theoretical studies of 1D bosons goes back to the celebrated work by Leib and Liniger, who found exact integrability of
zero-range interacting bosons via the Bethe ansatz, for all values of the interaction strength \cite{BetheAnsatzExactSolution}.
There are two limiting cases for 1D systems: the high-density weak-interaction Thomas-Fermi (TF) regime \cite{ThomasFermi}, where
the Bogoliubov energy functional \cite{Bogoliubov} and the thermodynamic limit of the Gross-Pitaevskii mean-field theory
\cite{GrossPitaevskii} apply; and the low-density strong interaction Tonks-Girardeau (TG) regime of impenetrable bosons
\cite{TonksGirardeauRegime,Girardeau}.

Unlike the TF regime, in the TG regime the boson wavefunction is Fermi-like and the fermion-boson duality method have been
proposed for this regime \cite{Girardeau}. The idea of this method is to substitute dynamical correlations due to strong two-body
interactions by statistical correlations due to "Fermi" statistics of impenetrable bosons in 1D
\cite{FermionBosonDuality,FermionizedBosons}. For analytical studies, however, the original Leib-Liniger first-quantization
approach as well as the boson-fermion duality method are complicated since one has to solve the Schr\"odinger equation for the
many-particle wavefunction, which is a function of a macroscopically large number of variables $N$, the total number of particles
in the system. Therefore, it is desirable to possess a more handy method, {\it e.g.}, a method based on path-integral concepts.

In this letter we propose such a path-integral approach for the exact transformation between 1D bosons and 1D fermions. The
method is actually a 1D analogue of the 2D composite particles formalism, which enables mapping fermions on bosons and vise versa
in 2D by the coupling to the Chern-Simons gauge field \cite{Lopez}. As a result, we obtain for the first time the
long-wave-length asymptotics of a collective mode in the TG regime. The mode turns out to be of a Bogoliubov form.

We start from a secondary quantized Hamiltonian of a homogeneous system of spinless 1D bosons, which interact through a two-body
potential $U$:
\begin{eqnarray*}
\hat H &=&\hat K+\hat U,\\
\hat K &=& \frac1{2m}\int dx  \hat\psi_b^\dagger(x) (-i \partial_x)^2 \hat\psi_b(x), \\
\hat U&=&\frac 12 \int dxdx' \delta\hat\rho_b(x) U(x-x')\delta \hat\rho_b(x'),
\end{eqnarray*}
where $m$ is the mass; $\partial_x=\partial/\partial x$; $\hat\rho_b(x)=\hat\psi_b^\dagger\hat\psi_b(x)$ with $\hat\psi_b$ being
the boson operators and $\delta \hat \rho_b(x)\equiv \hat\rho_b(x)-\bar\rho_b$ is the density fluctuations where $\bar\rho_b$ is
the average boson spatial density.

The many-particle wavefunction in the TF regime experiences no crucial changes as one particle passes another, whereas in the TG
regime it falls down \emph{almost} to zero as the coordinate of one particle approaches the position of another one. The
"fermionized" boson wavefunction has \emph{zero} value if the position of one particle coincides with that of another one. Such a
reduction of the Hilbert space to fermionized wavefunctions is an approximation, but is justified far inside the TG regime.

A fermionized boson wavefunction can be constructed from a fermion antisymmetric wavefunction in the following way
\cite{Girardeau,FermionBosonDuality}:
\begin{eqnarray}
\psi_b\left(\left\{x_i\right\}\right) =
\prod\limits_{i<j}\text{sign}\left(x_i-x_j\right)\psi_f\left(\left\{x_i\right\}\right)\label{FBfunction}.
\end{eqnarray}

In a secondary-quantized language Eq. (\ref{FBfunction}) corresponds to the introduction of the new quasi-particle operators
$\hat\psi_f^\dagger$, which are related to $\hat\psi_b^\dagger$ by
\begin{eqnarray}
\hat\psi_b^\dagger(x) = \hat\psi_f^\dagger(x)\exp\left(-i\pi\int dx'\theta(x-x')\hat\rho(x')\right),\label{transformation}
\end{eqnarray}
where $\theta$ is the Heaviside unit step function and
\begin{eqnarray*} \hat\rho(x)\equiv\hat\psi_b^\dagger(x)\hat\psi_b(x) =
\hat\psi_f^\dagger(x)\hat\psi_f(x)
\end{eqnarray*}
is the spatial particle density, which has the same form in terms of initial bosons and the new quasi-particles. It is easy to
prove that
\begin{eqnarray*}
\hat\psi_f(x_1)\hat\psi_f^\dagger(x_2)-e^{-i\pi\Delta}\hat\psi_f^\dagger(x_2)\hat\psi_f(x_1)\\
=\left\{\hat\psi_f(x_1),\hat\psi_f^\dagger(x_2)\right\}=\delta(x_1-x_2),
\end{eqnarray*}
where $\Delta = \theta(x_1-x_2)-\theta(x_2-x_1)=\text{sign} (x_1-x_2)=\pm1$. That is, the new operators satisfy Fermi
anti-commutation relations so that the quasi-particles are fermions. Let us call them composite fermions (CF), after their
predecessors in 2D.

To see that the transformation (\ref{transformation}) corresponds to Eq.(\ref{FBfunction}), note that if one starts creating
boson wavefunction by repeatedly acting on a vacuum state with the operators $\hat\psi_b^\dagger$ from Eq.(\ref{transformation}),
then the CF operators will produce fermionic wavefunction and the exponential phase-factors will give the "$\text{sign}$" term in
Eq.(\ref{FBfunction}).

The kinetic energy and two-body interaction in the CF operators' representation take the following forms
\begin{eqnarray*}
\hat K &=& \frac 1{2m}\int dx \hat\psi^\dagger_f(x)(-i\partial_x + k_F + \hat a_x)^2\hat\psi_f(x),\\
\hat U &=& \frac 1{2\pi^2} \int dx dx' \hat a_x(x) U(x-x') \hat a_x(x'),
\end{eqnarray*}
with the constraint
\begin{eqnarray}
\hat a_x(x) = \pi \delta \hat\rho(x), \label{Constraint}
\end{eqnarray}
where $k_F=\pi \bar\rho$ is the Fermi wave-vector.

In a path-integral representation the constraint (\ref{Constraint}) is easily incorporated with the aid of a Lagrange multiplier.
The partition function has the following form so far:
\begin{eqnarray}
{\cal Z}(\phi) = \int {\cal D} \psi_f{\cal D} \psi^\dagger_f{\cal D}a_x e^{i\int dt {\cal L}}
\prod\limits_{\{t,x\}}\delta\left(\frac{a_x}\pi - \rho + \bar\rho\right),
\label{constainedintegration}\\
\nonumber {\cal L} = \int dx \left(\psi^*_f(i\partial_t)\psi_f+\rho \phi \right) - K(\psi^*_f,\psi_f) - U(a_x),
\end{eqnarray}
where $\rho\equiv \psi^*_f\psi_f$ and the constrained path integration is over the statistical field $a_x$ and the Grassmann
fields $\psi_f$ and $\psi^*_f$ which represent the CFs. An external potential $\phi$ has also been added to the action. We will
use it to probe the system, {\it i.e.}, the density-density correlation function is:
\begin{eqnarray}
\left\langle\hat\rho(t,x)\hat\rho(t',x')\right\rangle =-\left.{\cal Z}(\phi)^{-1}\frac{\delta^2{\cal
Z(\phi)}}{\delta\phi(t,x)\delta\phi(t',x')}\right|_{\phi=0}.\label{CorrelatorDefinition}
\end{eqnarray}

The constraint in the partition function (\ref{constainedintegration}) can be rewritten through the introduction of an auxiliary
field $a_t$ as:
\begin{eqnarray*}
\prod\limits_{\{t,x\}}\delta\left(\frac{a_t}\pi - \rho + \bar\rho\right)= \int {\cal D} a_t(t,x) e^{-i\int dt dx
\left(\frac{a_x}\pi - \rho + \bar\rho\right)a_t}.
\end{eqnarray*}
As a result, the spatial density of the Lagrangian becomes
\begin{widetext}\begin{eqnarray*}
{\cal L}&=& \psi^*_f\left( i\partial_t - \frac {(-i\partial_x + k_F + a_x)^2}{2m}\right)\psi_f  - a_t\left(\frac{a_x}\pi - \rho +
\bar\rho\right) +\rho \phi - \frac 1 {2\pi ^2} \int dx' a_x(x) U(x-x') a_x(x'),
\end{eqnarray*}\end{widetext}
and the integration in ${\cal Z}$ is assumed now over a time-space statistical field $\bm{a}=(a_t,a_x)$.

Before proceeding further, let us outline the connection of the proposed boson-fermion transformation to existing theories. The
transformation (\ref{transformation}) is reminiscent of the inverse boson-fermion transformation in Haldane's bosonization
approach for 1D Fermi liquids \cite{Bozonization}. Nevertheless, Haldane's bosonization is developed for studies of low-frequency
physics of fermion systems in terms of bosons, which, in fact, represent the spatial density fluctuations of the fermions. In our
case, however, the transformation (\ref{transformation}) involves operators of real bosons and not density fluctuations. Density
fluctuations, instead, are represented by the spatial statistical field component $a_x$.

Physically, the proposed model is a 1D counterpart of Jain's mechanism of attaching flux quanta to 2D particles, which leads to
the coupling of the composite objects to a Chern-Simons gauge field \cite{Lopez}. There are, however, several aspects in which
the theory proposed differs from the Chern-Simons theory, apart from the different dimensionalities of the systems. ({\it i}) The
proposed theory is not gauge invariant. ({\it ii}) Time-reversal and space-reversal symmetries are broken separately, though
time-space-reversibility is present. ({\it iii}) The coupling to Chern-Simons gauge fields in 2D results in an additional
magnetic field experienced by the composite objects, whereas in our case the coupling to the statistical fields leads to a shift
of the one-particle kinetic energy dispersion by $k_F$ in momentum space. In a sense, it looks as though the whole system starts
moving. This fact is going to reveal itself later through a Doppler shift in the response function.

The noninteracting part of the action becomes the sum of two Gaussian actions for the CFs and the statistical fields governed by
the following "bare" CFs' and statistical fields' propagators respectively (in Fourier space):
\begin{eqnarray*}
G_0^{-1}(\bm{p}) &=& \varepsilon - \frac{(p-k_F)^2}{2m},\\
\hat D_0^{-1}(\bm{q})&=&-\frac1{\pi} \left[\begin{array}{cc}0&1\\1&v_F u(q)\end{array}\right],
\end{eqnarray*}
with $u(q)\equiv U(q)/(v_F\pi)$, where $U(q)$ is the spatial Fourier transform of the two-body potential and
$\bm{p}=(\varepsilon,p), \bm{q}=(\omega,q)$. The interaction part of the action is:
\begin{eqnarray*}
{\cal L}_{int} &=& (\phi + a_t) \rho - a_x j,
\end{eqnarray*}
where $j$ is the CF current density:
\begin{eqnarray*}
j = \frac1{2m}(\psi^*_f((-i\partial_x-k_F)\psi_f) + ((i\partial_x-k_F)\psi^*_f)\psi_f + a_x\rho).
\end{eqnarray*}
Due to the coupling to the statistical fields the one-particle dispersion is shifted by $k_F$ in momentum space. One can formally
make a substitution $p-k_f\to p$ and arrive at the ordinary picture of 1D fermions at rest.
\begin{figure}[tb]
\includegraphics[width=8.7cm]{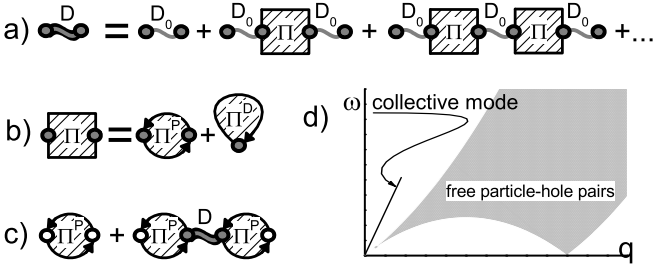}
\caption{\label{Figure1} (a)-(c) Graphical representation of Eqs.(\ref{TotalPropagator}-\ref{DensityResponse}). The hollow
interaction vertices in (c) denote the density components of the two-component density-current interaction vertex. (d) The
excitation spectrum of a 1D boson system at the TG regime. Besides free particle-hole pair excitations there is a collective mode
with a Bogoliubov-like dispersion $\omega(q)=q\sqrt{\bar\rho U(q)/m}$.}
\end{figure}

As the action is Gaussian in the CF fields, one can integrate them out. The integration leads as usual to the fermion determinant
in the effective statistical fields' action. At this point it is important to note that so far we made no approximations. Now in
order to obtain the density-density response of the system it suffices to leave in the effective action only the terms quadratic
in the statistical fields:
\begin{eqnarray*}
{\cal Z}_{\text{eff}}(\phi)&=&\int {\cal D}\tilde {\bm{a}} \text{exp}\left(i \frac12\int\frac{d^2 \bm{q}}{(2\pi)^2}{\cal
L}_{\text{eff}}\right).
\end{eqnarray*}
Here, the effective action has the following form
\begin{eqnarray*}
{\cal L}_{\text{eff}}&=& \bm{a} (-\bm{q})(\hat D_0^{-1}(\bm{q})-\hat \Pi^D(\bm{q})) \bm{a}(\bm{q})\\
&&-(\bm{a}+\phi)(-\bm{q})\hat \Pi^P(q)(\bm{a}+\phi)(\bm{q}),
\end{eqnarray*}
where $\Pi^D$ and $\Pi^P$ are the diamagnetic and paramagnetic polarization operators respectively. The renormalized statistical
fields' propagator is given by a Dyson equation:
\begin{eqnarray}
\hat D(\bm{q}) &=& \left(\hat D^{-1}_0(\bm{q})-\hat \Pi(\bm{q}))\right)^{-1},\label{TotalPropagator}
\end{eqnarray}
where
\begin{eqnarray}
\hat\Pi(\bm{q})&=&\hat\Pi^P(\bm{q})+\hat\Pi^D(\bm{q})\label{TotalPolarization}
\end{eqnarray}
is the total polarization operator. Using Eq.(\ref{CorrelatorDefinition}) one can now find the density-density response in the
following form (see Fig.\ref{Figure1}):
\begin{eqnarray}
\left\langle\rho(-\bm{q})\rho(\bm{q})\right\rangle=[\hat \Pi^P(\bm{q})+\hat \Pi^P(\bm{q})\hat D(\bm{q})\hat
\Pi^P(\bm{q})]_{tt}\label{DensityResponse} .
\end{eqnarray}

The advantage of the proposed transformation to CFs is that in normal Fermi liquids \footnote{We assume that the system of the
CFs belongs to the class of Fermi liquids with strong forward scattering. The BCS and CDW instabilities are unlikely to appear in
our case because they would have produced a gap in the excitation spectrum, which the initial boson system can not possess.} we
know the exact form of the polarization operator in the long-wave-length limit \cite{Voit}. Due to the so-called
loop-cancellation theorem (see, {\it e.g.}, Ref.\cite{DiCastro}), in the long-wave-length limit the polarization operator is
given by an RPA expression. In fact, the loop-cancellation theorem is a consequence of gauge invariance, so that one can argue
that we can not use it in our case. However, far inside the TG regime (see below) the gauge invariance is restored so that the
theorem still applies. In 1D one has ($q\ll k_F$)
\begin{eqnarray*}
\hat \Pi^D(\bm{q})&=&\frac1{\pi}\left[\begin{array}{cc}0&0\\0&v_F\end{array}\right],\\
\hat \Pi^P(\bm{q})&=&\frac1{\pi}\frac1{\omega'^2-(1-i0^+)^2}\left[\begin{array}{cc}v_F^{-1}&-\omega'\\-\omega'&
v_F\end{array}\right],
\end{eqnarray*}
with $\omega'=\omega/(v_F q)$. Substituting these expressions into Eq.(\ref{DensityResponse}) one gets the density-density
response in the long-wave-length limit:
\begin{eqnarray*}
\left\langle\rho(-\bm{q})\rho(\bm{q})\right\rangle=\frac{1}{\pi v_F((\omega'-1)^2-(u(q)+2-i0^+))}.
\end{eqnarray*}

As mentioned, we obtained the Doppler shift since the proposed theory lacks time-reversal symmetry. Nevertheless, as one goes
toward the strong TG regime
\begin{eqnarray}
u(q)\to\infty,\label{condition}
\end{eqnarray}
the time-reversal symmetry is restored leading to the following density-density response:
\begin{eqnarray*}
\left\langle\rho(-\bm{q})\rho(\bm{q})\right\rangle_b=\frac{v_F}{\pi}\frac{q^2}{\omega^2-(v(q) q - i0^+)^2},
\end{eqnarray*}
where the momentum dependent "velocity" of the collective mode is:
\begin{eqnarray}
v(q)=\sqrt{\bar\rho U(q)/m}.\label{velocity}
\end{eqnarray}

The fact that our theory gives reasonable results only far inside the TG regime is natural. The fermionization of bosonic
wavefunctions (\ref{FBfunction}) is a good approximation if, and only if, the inter-boson repulsion is very strong, {\it i.e.},
if the condition (\ref{condition}) is satisfied.

The excitation spectrum of 1D bosons in TG regime (see Fig.\ref{Figure1}d) now appears to consist of two kinds of excitations,
just like in Fermi liquids. The first one, originally discussed by Lieb \cite{Lieb}, is free particle-hole pairs. The dispersion
of these excitations is given by the density of states of the polarization operator $\Pi$. In the long-wave-length limit they are
phonons with velocity $v_F$. The second one is the collective mode with velocity (\ref{velocity}). This collective mode is a
surprising result because it has the form of Bogoliubov phonons, and in the TG regime the Bogoliubov theory is not applicable.

Another issue is the effective two-body interaction, through which the CFs interact. In the limit (\ref{condition}) the gauge
invariance is restored and we are left only with density-density interactions in the system, because only the time-time component
of the renormalized statistical field propagator survives:
\begin{eqnarray*}
D_{tt}(\bm{q}) = \pi v_F u(q)(\omega'^2-1)/(\omega'^2-u(q)).
\end{eqnarray*}

In the static limit ($\omega'\to 0$) the effective interaction between CFs is $D_{tt} = \pi v_F$. This reflects the well-know
fact that unlike the initial boson system, in the fermionized case the two-body interaction becomes weak. Notice, however, that
in our case with the increase in initial inter-boson potential, $u(q)$, the effective two-body potential between the CFs does not
vanish, as in Ref.\cite{FermionBosonDuality}, but instead reaches a constant value.

In the limit of high energy transfer, $\omega' \gg u(q)$, we have $D_{tt}(\bm{q}) = U(q)$. This is natural, since at high
energies the screening is relatively weak so that the particles interact via the initial potential $U(q)$.

Finally, in the intermediate region, $1<\omega'<u(q)$, the so-called overscreening effect takes place and the effective potential
becomes attractive. This fact may have interesting consequences.

In conclusion, we propose a path-integral method for statistics transformation in 1D, which leads to the coupling of composite
particles to statistical fields, which are representatives of density-current interactions. With the aid of this transformation
we have found a long-wave-length limit of a collective mode in TG regime of the boson system. The method proposed, however, can
be used for further study of TG bosons, {\it e.g.}, the study of response of the system near $q\sim 2k_F$, the question of the
CDW and BCS stabilities, the effect of statistical field fluctuations on CF propagator {\it etc.} We leave these investigations
for a later work.

This work was supported by the NSF and the PRF.

\end{document}